\documentclass[sigconf]{acmart}

\AtBeginDocument{%
  }
\setcopyright{acmlicensed}
\copyrightyear{2024}
\acmYear{2024}
\acmDOI{XXXXXXX.XXXXXXX}

\acmConference[Conference'24]{TBA Conference}{2024}{USA}
\usepackage[T1]{fontenc}
\usepackage{epsfig,endnotes}
\usepackage{import}
\usepackage{url}
\usepackage{commath}
\usepackage{booktabs}  
\usepackage{amsfonts}     
\usepackage{nicefrac}     
\usepackage{microtype} 
\usepackage{multirow}
\usepackage{graphicx}
\usepackage{xcolor}
\usepackage{subcaption}
\usepackage{array}
\usepackage[linesnumbered,ruled,vlined]{algorithm2e}
\usepackage{pifont}
\usepackage{soul}
\usepackage{epigraph}

\newcommand{\cmark}{\ding{51}}%
\newcommand{\xmark}{\ding{55}}%

\def\name{TATTOOED\xspace}

\SetKwInput{KwInput}{Input}             
\SetKwInput{KwOutput}{Output}  

\begin{document}

\title{\name: A Robust Deep Neural Network Watermarking Scheme based on Spread-Spectrum Channel Coding}

\author{Giulio Pagnotta}
\affiliation{%
 \institution{Sapienza University of Rome}
  \country{Italy}
}
\orcid{0000-0002-4626-6045}
\email{pagnotta@di.uniroma1.it}

\author{Dorjan Hitaj}
\affiliation{%
 \institution{Sapienza University of Rome}
  \country{Italy}
}
\orcid{0000-0001-5686-3831}
\email{hitaj.d@di.uniroma1.it}

\author{Briland Hitaj}
\affiliation{%
 \institution{SRI International}
 \country{USA}
}
\orcid{0000-0001-5925-3027}
\email{briland.hitaj@sri.com}

\author{Fernando Perez-Cruz}
\affiliation{%
 \institution{Swiss Data Science center}
  \country{Switzerland}
}
\orcid{0000-0001-8996-5076}
\email{fernando.perezcruz@sdsc.ethz.ch}

\author{Luigi V. Mancini}
\affiliation{%
 \institution{Sapienza University of Rome}
  \country{Italy}
}
\orcid{0000-0003-4859-2191}
\email{mancini@di.uniroma1.it}

\renewcommand{\shortauthors}{Pagnotta et al.}

\begin{abstract}

Watermarking of deep neural networks (DNNs) has gained significant traction in recent years, with numerous (watermarking) strategies being proposed as mechanisms that can help verify the ownership of a DNN in scenarios where these models are obtained without the permission of the owner. However, a growing body of work has demonstrated that existing watermarking mechanisms are highly susceptible to removal techniques, such as fine tuning, parameter pruning, or shuffling.

In this paper, we build upon extensive prior work on covert (military) communication and propose \name, a novel DNN watermarking technique that is robust to existing threats. We demonstrate that using \name as their watermarking mechanisms, the DNN owner can successfully obtain the watermark and verify model ownership even in scenarios where $99\%$ of model parameters are altered. Furthermore, we show that \name is easy to employ in training pipelines, and has negligible impact on model performance. 

\end{abstract}

\keywords{DNN watermarking, channel coding, IP protection}

\ccsdesc[500]{Computing methodologies~Machine learning}
\ccsdesc[500]{Security and privacy}
\settopmatter{printfolios=true}
\maketitle

\section{Introduction}\label{sec:introduction}

The presence of vast quantities of data originating from multitudes of sources in conjunction with powerful computational resources has fueled the last decade of machine learning (ML) applications. In particular, deep learning (DL) has demonstrated exceptional results in a growing number of domains, constantly pushing the boundaries of previously known state-of-the-art solutions. The abundance of significant volumes of data enables DNNs, the core ML technique at the heart of many state-of-the-art DL solutions, to autonomously determine and learn relevant features directly from the input, removing the need for handcrafted features. In doing so, DNNs have transformed themselves as the ML-technique of choice for areas such as image classification~\cite{Simonyan14verydeep,He2016DeepRL,Chollet2017XceptionDL}, natural language processing~\cite{nlp1,nlp2}, speech recognition~\cite{speech1,speech2}, data (image, text, audio) generation~\cite{styleGAN,passGAN,iresnet,9833616}, and cyber-security~\cite{encod,de2022evading,piskozub2021malphase,pagnotta2023dolos,de2024have}.

However, despite their undeniable advantages, training good DL models comes with a series of challenges: 1) While frameworks such as TensorFlow or PyTorch make it easy to construct DL pipelines for a target problem, identifying the right DNN architecture for the task (i.e., the correct number of layers or hyper-parameters to use) can be a challenging task even for ML experts. 2) The absence of required computational resources can make it difficult for an entity to benefit from DL. 3) Large quantities of training data can incorporate proprietary characteristics, thus requiring additional layers of protection when the resulting models are made available to the public, including machine learning as a service (MLaaS). 
Therefore, companies, institutions, and even individuals devising and training \emph{proprietary} DNN models want to protect this new form of intellectual property (IP) from the prying eyes of competition and malicious adversaries.
There has been a growing body of work aiming at devising mechanisms that would enable an entity to verify, with high assurance, the legitimate ownership of a suspected DNN model. In particular, \textbf{DNN watermarking}, a concept first introduced by Uchida et al.~\cite{uchida_embedding} has gauged significant interest in the research community, resulting in a surge of watermarking techniques~\cite{wang_dnn2020,song2017machine,deepsign2019,deepmark_2019,backdoor2018watermark,Chen2019BlackMarksBM,Jia2021EntangledWA,Merrer2017AdversarialFS,Yang2019EffectivenessOD,namba2019robust,Szyller2021DAWN}. However, as it commonly happens in security and privacy works, this has led to the development of novel methods demonstrating that it is possible to remove~\cite{chen_refit,Hitaj2019evasion,wang2019attacks,namba2019robust,neural_cleanse}, overwrite~\cite{wang2019attacks}, and forge~\cite{Xu2019ANM,fan2019rethinking} the watermarks placed in a DNN. Thus, the development of a secure DNN watermarking strategy remains an open problem.

\begin{figure*}[htp]
    \centering        
	    \begin{subfigure}{.45\textwidth}
            \centering
            \includegraphics[width=\columnwidth]{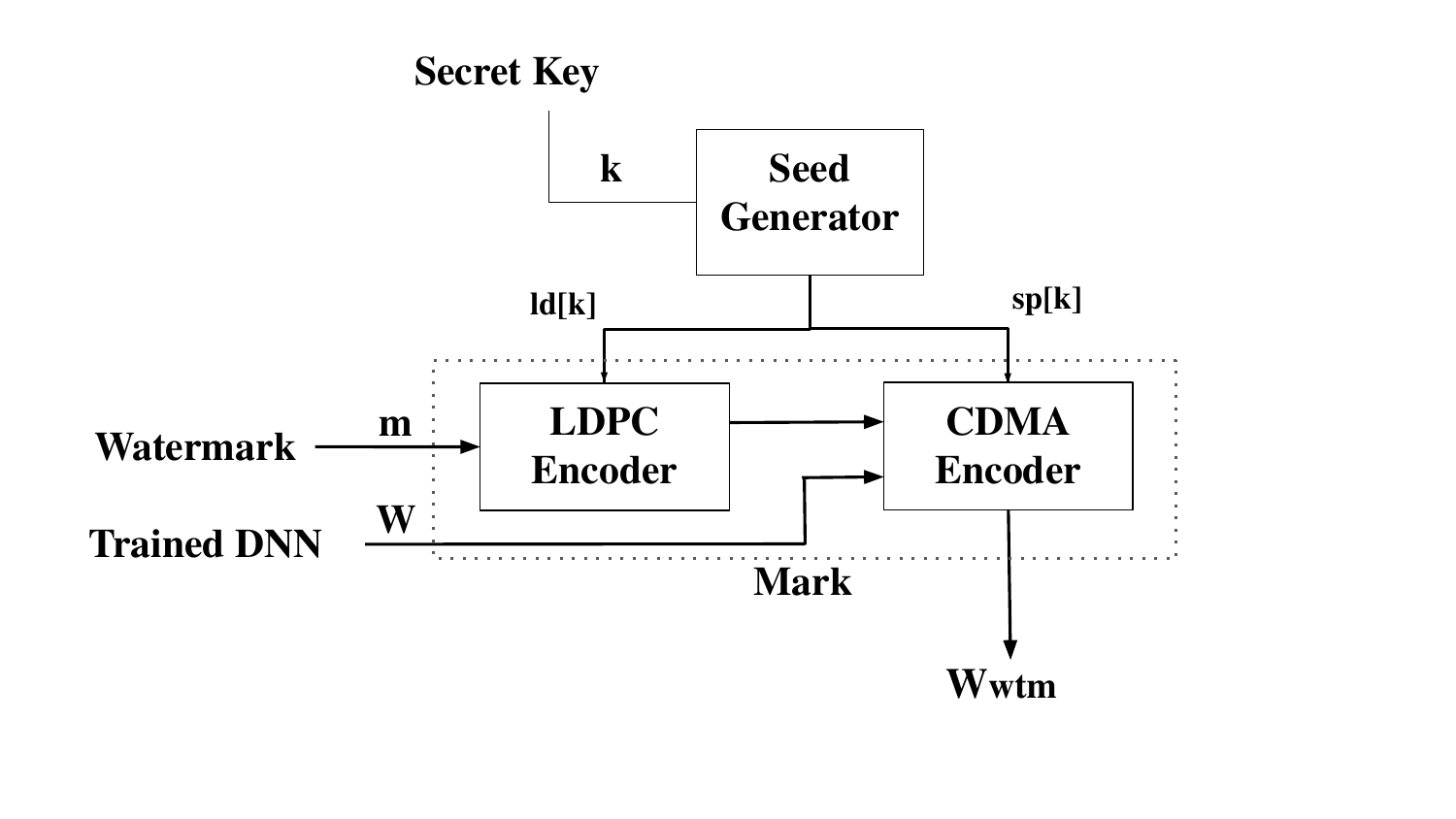}
            
             \caption{\name watermark embedding procedure.}
             \label{fig:scheme_mark}
        \end{subfigure}
        \hfill
	    \begin{subfigure}{.45\textwidth}
            \centering
            \includegraphics[width=\columnwidth]{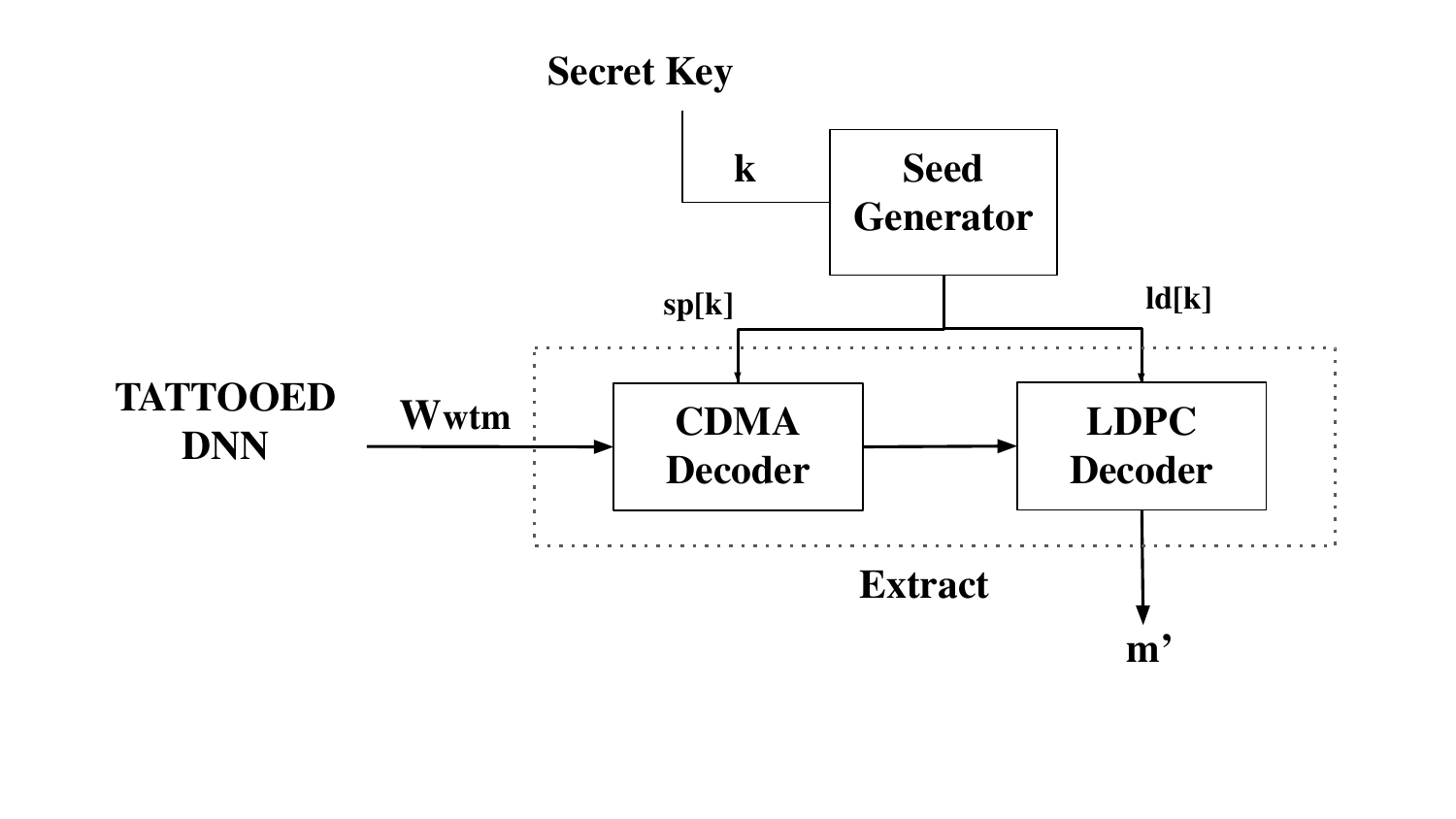}
             \caption{\name watermark extraction procedure.}
             \label{fig:scheme_verify}
        \end{subfigure}
         \bigskip
    \vspace{-1em}
    \caption{A high-level overview of \name watermarking scheme. To watermark a DNN (\ref{fig:scheme_mark}), \name uses a secret key \textbf{k}, out of which two separate keys (\textbf{ld[k]} and \textbf{sp[k]}) are generated. \textbf{ld[k]} is used in the LDPC encoder (and decoder on extraction procedure) and \textbf{sp[k]} is used to generate the spreading codes (both in CDMA encoder and decoder). The LDPC encoder takes the watermark \textbf{k}, encodes it, and then passes it to the CDMA encoder, which spreads the LDPC-encoded watermark in a wider bandwidth and then adds it the selected parameters of model $W$ to obtain the watermarked model $W_{wtm}$. To extract the watermark (\ref{fig:scheme_verify}), a watermarked model using \name is given as input to the \textbf{Extract} module, passing first to the CDMA decoder, and then to the LDPC decoder to recover the hidden watermark \textbf{m'}.}
    \label{fig:scheme_tattooed}
\end{figure*}

This paper introduces \name, a novel 
watermarking framework for deep neural networks that pushes forward the state-of-the-art in DNN watermarking. The proposed framework uses Code Division Multiple Access (CDMA) spread-spectrum channel coding to embed a watermark directly into the weight parameters of a DNN model. 
This unique approach is highly robust and secure against existing state-of-the-art watermark removal techniques~\cite{model_compression,shafieinejad_watermarking_robustness,chen_refit}. To the best of our knowledge, this is the first time CDMA has been used in DNN watermarking, making \name a pioneering work in the field.
Our innovation also includes the technique of applying CDMA to watermark a DNN model. We use CDMA to insert a watermark directly into the weight parameters of the DNN model, causing minimal changes to the model's parameters. This allows \name to watermark a DNN model without impacting its performance on the intended task. Additionally, our method of embedding the watermark directly into the model's parameters makes \name independent of the DNN architecture, which means it can be applied to different DNN models. This is a major advancement in the field of DNN watermarking as it allows for more flexibility and adaptability.
Furthermore, the \name watermark embedding method does not necessitate any modifications to the supply chain used to create a DNN model, as the watermark can be added at the conclusion of standard training. This makes \name easy to adopt with minimal technological impact, even when producing complex and costly Large Language Models (LLMs) models such as GPT~\cite{NEURIPS2020_1457c0d6} or Llama\cite{touvron2023llama}. This is a significant novelty compared to previous DNN watermarking methods that embed secret bits during model training and require supply chain customization. 
\name ability to add a watermark post-training and without changes to the supply chain gives it a clear advantage over existing methods in terms of ease of adoption and low overhead.

Our contributions can be summarized as follows:
\begin{itemize}
    \item We introduce \name, a novel deep neural network watermarking technique based on CDMA spread-spectrum channel-coding, which is the first of its kind in the field of DNN watermarking.
    \item We show that \name is secure against state-of-the-art watermark removal techniques~\cite{model_compression,shafieinejad_watermarking_robustness,chen_refit} and demonstrate that the \name watermark cannot be removed. Our thorough evaluation demonstrates that \name is more secure than prior work on DNN watermarking, thus setting a new benchmark for DNN watermarking security and robustness. 
    \item We show that \name is a general-purpose scheme for DNN watermarking through an extensive empirical evaluation under varying conditions: a) multiple domains, including image, text, and audio; b) different DNN architectures; c) different classification tasks; d) several benchmark datasets, and e) diverse watermark sizes. 

\end{itemize}

\section{Background}\label{sec:background}

\subsection{DNN Model Watermarking}\label{sec:background_dnnwatermarking}
Digital watermarking was introduced to covertly embed a secret message, the watermark, inside digital data (e.g., an image) to provide proof of ownership of the watermarked data. There are two classes of watermarking algorithms: \textit{zero-bit} watermarking, where the watermark extraction corresponds to a detection task, and \textit{multi-bit} watermarking, where the message corresponds to a $P$-bits watermark $\mathbf{m}$. 
Multi-bit watermarking has been shown to have superior flexibility with respect to zero-bit watermarking being able to be applied in a broader range of applications~\cite{multi_zero_comparison}.
Both zero-bit and multi-bit watermarking algorithms may require knowledge of the non-watermarked content to recover the watermark. In case they do not require any knowledge of the non-watermarked content, they are called \textit{blind} watermarking algorithms, and \textit{non-blind} otherwise~\cite{barni_survey_wtm}.

Digital watermarking can be applied to DNNs. To prove the ownership of a model $\mathbf{W}$, the owner needs a way to embed a secret message inside the model parameters $\mathbf{w} \subseteq \mathbf{W}$, and a way to extract the message reliably. A DNN watermarking scheme consists of a message space $\mathbb{M}$, a key space $\mathbf{K}$, and the following \textbf{Mark} and \textbf{Extract} algorithms: 
\textbf{Mark} is a polynomial and deterministic algorithm that given a secret key $k \in \mathbb{K}$, a model $\mathbf{W}$, and a watermark message $\mathbf{m} \in \mathbb{M}$, outputs a model $\mathbf{W}_{wtm}$ that contains the \textit{watermark} $\mathbf{m}$.  
\textbf{Extract} is a polynomial and deterministic algorithm that, given a secret key $k \in \mathbb{K}$ and a model $\mathbf{W}$, outputs a $P$-bits sequence $\mathbf{m'}$ extracted from the model $\mathbf{W}$. 

\leavevmode \\
\textbf{Watermarking Requirements.}
Prior works in the domain have outlined several conditions for a DNN watermarking scheme to succeed~\cite{backdoor2018watermark,deepmark_2019,barni_survey_wtm}. Most of the scientific literature considers several fundamental requirements that a DNN watermarking scheme should satisfy. Below we describe each of these requirements, which also correspond to those we aim to fulfil in our work.
\begin{itemize}
    \item \textbf{Fidelity} requirement guarantees that the \textbf{Mark} algorithm does not significantly impair the performance of the model.
    
    \item \textbf{Robustness} requirement guarantees that the \textbf{Mark} algorithm embeds the watermark in such a way that it cannot be disrupted by the most common manipulations that the watermarked model may undergo during its life cycle.

    \item \textbf{Security} requirement guarantees that the \textbf{Mark} algorithm embeds the watermark securely, i.e., the watermark cannot be disrupted by malicious manipulations of the watermarked DNN model. A DNN watermarking algorithm is secure if the malicious entity to destroy the watermark: a) either renders the model unusable in its intended task or b) requires a cost, in terms of resources (e.g., data, time, computational power), greater than the cost that the legitimate owner incurred to produce the watermarked model.
    
    \item \textbf{Integrity} requirement guarantees that, in absence of modifications to the watermarked model, the \textbf{Extract} algorithm outputs the watermark message $\mathbf{m'}$ equal to the originally inserted watermark message $\mathbf{m}$. 

    \item \textbf{Efficiency} requirement guarantees that the \textbf{Mark} and \textbf{Extract} algorithms can efficiently embed the watermark and verify its presence in a model without incurring a large computational overhead.

    \item \textbf{Generality} requirement guarantees that the \textbf{Mark} and \textbf{Extract} algorithms can be used on every kind of model architecture and any learning task.

\end{itemize}

\leavevmode \\
\textbf{Watermark Performance Metrics.} 
The performance of a DNN watermarking scheme is measured on two equally-important distinct aspects. Those performance parameters are the Test Error Rate (TER) and the watermark Bit Error Rate (BER).
TER measures the DNN model's performance and is calculated as ($1-ModelAccuracy$). A good watermarking scheme should cause minimal change to the TER of the model post-watermarking compared to the TER of the model before watermarking. 
On the other hand, BER measures the watermark extraction performance by calculating the rate of the erroneously extracted watermark bits, thus displaying the watermark accuracy. 
In Section~\ref{sec:evaluation} we evaluate the performance of \name watermarking scheme based on these two performance metrics.

\subsection{Spread-Spectrum Channel Coding}\label{sec:cdma_background}
In digital communications, spread-spectrum techniques~\cite{spread_spectrum_principles} are methods by which an electrical, electromagnetic, or acoustic signal with a particular bandwidth is deliberately spread in the frequency domain. These techniques enable spreading a narrowband information signal over a wider bandwidth. The receiver, knowing the spreading mechanism, can recover the original bandwidth signal upon receiving.
Two main techniques are used to spread the bandwidth of a signal: \textit{frequency hopping} and \textit{direct sequence}. In \textit{frequency hopping}, the narrowband signal is transmitted for a few seconds in a given band that is constantly changed using a pseudo-random frequency band that has been agreed with the receiver. The receiver, in coordination, tunes its filter to the agreed-on frequency band to recover the message. 
\textit{Direct Sequence}, works by directly coding the data at a higher frequency using pseudo-random generated codes that the receiver knows. CDMA, the technique we employ in \name, is part of direct sequence channel coding techniques.
In CDMA, the transmitters use pseudo-random codes to encode their data making the spread-spectrum appears random and having noise-like properties. This guarantees privacy because the receiver cannot demodulate the transmitted signal without knowing the pseudo-random sequence or the method to generate the sequence used to encode the data. Another crucial property of CDMA is its resistance to jamming. The wide bandwidths that CDMA can achieve require the jamming signal to have a large amount of power, which is typically not feasible in practice. As we show in this paper, the inherent security properties of CDMA and its ability to survive malicious disruptive attacks make CDMA an excellent choice to be employed in DNN watermarking.

In \name we rely on CDMA to embed the watermark in the weights of the model. The intuition of behind \name watermark embedding is the following: each bit of the watermark is spread using a different CDMA spreading code, so \name works as if each bit of the watermark is being transmitted/embedded at the same time by a different user. The DNN weights chosen for watermarking can be seen as the noise in the communication channel. The receiver (which, in the \name case, is the watermark verifier) can use the individual spreading codes of each bit to decode the complete watermark. We explain how \name works in detail in Section~\ref{sec:tattoed}.
\vspace{-1em}
\section{Threat Model}\label{sec:threat_model}

Our threat model considers the design and development of a DNN watermarking technique that can be used for both ownership verification and traitor tracing~\cite{traitor_tracing}. In this way, the legitimate owner, besides verifying ownership of a DNN model, can also trace it back to an individual who has illegally redistributed the model. 
\leavevmode \\
\textbf{Adversary's Capabilities}.
The adversary corresponds to an entity that has gained illegitimate access to a proprietary ML model. The adversary is aware of the possibility of the model being watermarked and employs existing state-of-the-art watermark-removal techniques in an attempt to remove the watermark from the model.
The adversary wins if they can remove the watermark without incurring costs similar to what the legitimate owner spent to construct the model (e.g., the adversary could train a comparable model if he had to spend the same resources to remove the watermark).
\leavevmode \\
\textbf{Watermark verification.} Two ways exist to verify the presence of a watermark in a deep neural network. Those are white-box and black-box verification. In white-box watermark verification, the verifier (i.e., the legitimate owner) needs physical access to the model. In contrast, the verifier does not need such access in the black-box watermark verification. In a black-box scenario, the watermark verification is performed remotely by querying the model and observing the outputs returned by it.
We position ourselves in a white-box watermark verification scenario similar to prior works~\cite{Nagai2018DigitalWF,uchida_embedding,song2017machine,deepsign2019,deepmark_2019,wang_riga} in the field, where the verifier has white-box access to the DNN model to be verified.
We highlight that the white-box setting is the most used one for digital asset watermarking and intellectual property protection, and there seems to be a growing interest in the fully white-box watermark techniques~\cite{Nagai2018DigitalWF,uchida_embedding,song2017machine,deepsign2019,deepmark_2019,wang_riga}.

\section{TATTOOED}\label{sec:tattoed}
This section introduces \name, our \textit{white-box, blind, multi-bit} neural network watermarking technique. \name relies on applying the CDMA technique to embed a watermark in the weights of a deep neural network. The intuition of our watermark embedding is the following: each bit of the $\mathbf{m}$-bit watermark is spread using a different CDMA spreading code, so \name works as if each bit of the watermark is being transmitted at the same time by a different user. The DNN weights chosen for watermarking can be seen as the noise in the communication channel. The receiver (which, in the \name case, is the watermark verifier) can use the individual spreading codes of each bit to decode the complete watermark. This method of using CDMA makes the watermark stealthy and very robust against changes in DNN weights. 
In this scenario, a legitimate owner wants to insert a $P$-bits watermark $\mathbf{m}$ into the parameters of model $\mathbf{W}$. 
The bits of the watermark $\mathbf{m}$ are encoded as a vector of $\pm1$, and the code for each bit is represented by $\mathbf{c}_i$, which is a vector of $+1$ and $-1$ of length $R$, where $R$ is the number of parameters of model $\mathbf{W}$. $\mathbf{C}$ is an $R$ by $P$ matrix that collects all the codes. The matrix $\mathbf{C}$ is a secret known only by the legitimate owner of the DNN model, who owns the secret key $k$ that is used to generate $\mathbf{C}$. We assume the codes have been randomly generated with equal probabilities for $\pm 1$. 
The legitimate owner watermarks the model as follows: 
\begin{equation}
 \mathbf{W}_{wtm} = \mathbf{W} + \gamma \mathbf{C}\mathbf{m} \label{eq:embed_wtm}
 \end{equation}
where $\mathbf{W}$ denotes the vector containing the model parameters, and $\gamma$ is a parameter that affects the strength of the watermark signal. 
In Section~\ref{sec:experimental_setup}, we show how we determine the $\gamma$ parameter so that it does not impact the performance of the trained DNN model on the legitimate task. 
Now, the legitimate owner can recover the watermark hidden in the model's parameters using the spreading codes. For example, for each bit $i$ of the watermark, the legitimate owner can recover it as follows:
\begin{equation}
y_i = sign(\mathbf{c}_i^\top \mathbf{W}_{wtm})\label{eq:extract_wtm}
\end{equation} 
where $\mathbf{c}_i^\top$ is the transpose of $\mathbf{c}_i$.

The process of embedding and verifying the presence of the watermark in a DNN model is displayed by the pseudocodes of \textbf{Mark}, Algorithm~\ref{alg:mark_watermark} and \textbf{Extract}, Algorithm~\ref{alg:verify_watermark}.

\begin{algorithm}[htp]
\DontPrintSemicolon
\KwInput{Model: $W$, Int: $\gamma$, Str: $m$, Str: $key$, Float: $ratio$}
\KwOutput{Model: $W_{wtm}$}
    $spreading\_code\_seed, params\_seed \leftarrow seed\_gen(key)$ \;
    $R = len(W) * ratio$ \;
    $PRNG(params\_seed)$ \;
    $indices \leftarrow random([0, len(W)], size = R)$ \;
    $PRNG(spreading\_code\_seed)$ \;
    $i \leftarrow 0$ \;
    \While{$i < len(m)$} {
        $code \leftarrow random([-1, 1], size = R)$ \;
        $signal \leftarrow gamma * m[i] * code$ \;
        $W_{wtm}[indices] \leftarrow W[indices] + signal$  \;
        $i \leftarrow i + 1$ \;
  }
\caption{Mark}
\label{alg:mark_watermark}
\end{algorithm}

\textbf{Mark} (Algorithm~\ref{alg:mark_watermark}) takes as input a DNN model to be watermarked, the $\gamma$ value that will affect the power of the watermark signal, the encoded watermark content $\mathbf{m}$, a secret key, and the ratio parameter. 
The ratio parameter specifies the portion of the model's parameters $\mathbf{W} $ that will contain the watermark. In practice, to efficiently embed the watermark in network architectures with millions of parameters, it might be convenient not to use all of them to embed the watermark. The secret key consists of 512 bits and is used to generate separate secret seeds (line 1) that are individually used to generate the spreading code and to select the model parameter indices where the watermark will be embedded according to the ratio parameter. 
Line 2 computes the number of parameters $R$ selected to contain the watermark. On lines 3-4, we set the seed in the pseudorandom generator, and then we select $R$ indices from the vector $\mathbf{W}$ to embed the watermark. Then, we set the pseudorandom generator using the spreading code seed. Then, for each bit of the watermark $m$, a spreading code of the length of the number of the model parameters selected for watermarking is generated (line 8). The spreading code vector is multiplied by the bit value and the $\gamma$ (line 9). The resulting vector is added to the vector of the selected $R$ parameters of $W$ (line 10).

\begin{algorithm}[htp]
\DontPrintSemicolon
\KwInput{Model: $W$, Str: $key$, Float: $ratio$}
\KwOutput{Str: $m'$}
\KwData{Int: $wtm\_length$}
    $spreading\_code\_seed, params\_seed \leftarrow seed\_gen(key)$ \;
    $R = len(W) * ratio$ \;
    $PRNG(params\_seed)$ \;
    $indices \leftarrow random([0, len(W)], size = R)$ \;
    $PRNG(spreading\_code\_seed)$ \;
    $i \leftarrow 0$ \;
    $m' \leftarrow []$ \;
    \While{$i < wtm\_length$} {
        $code \leftarrow random([-1, 1], size = R)$ \;
        $y_i \leftarrow transpose(code) * W[indices]$  \;
        \If{$y_i > 0$} {
            $m'.append(0) $ \;
        }
        \Else {
            $m'.append(1) $ \;
        }
        $i \leftarrow i + 1$ \;
  }
\caption{Extract}
\label{alg:verify_watermark}
\end{algorithm}

\textbf{Extract} (Algorithm~\ref{alg:verify_watermark}) takes as input the model, the secret key, and the ratio parameter and returns the extracted watermark ($m'$). The necessary seeds are generated using the secret key (line 1). With such seeds, we select the model parameter indices needed to extract the watermark (lines 3-4) and generate the spreading codes (line 9). Note that the indices and spreading codes are the same used in \textbf{Mark} algorithm since they are generated using the same secret key $k$.
For each bit $i$ of the watermark $m'$, we retrieve a value $y_i$ by multiplying the transpose of the respective spreading code with the subset of the model parameters $W$ (line 10) and then assigning the bit 0 if the retrieved $y_i$ is greater than zero and the bit 1 otherwise (lines 9-14).

\section{Experimental Setup}\label{sec:experimental_setup}
\subsection{Datasets}

We evaluate \name on a series of benchmark datasets spanning domains like image, text, and audio recognition, see Table~\ref{tab:tattooed-datasets}.

\begin{table}[h]
\centering
\caption{The list of datasets, including their categories and description, used as part of \name evaluation.}
\label{tab:tattooed-datasets}
\begin{tabular}{|p{.6\columnwidth}|p{.3\columnwidth}|}
\hline

\textbf{MNIST}~\cite{mnist_dataset} & \emph{Image Recognition} \\ \hline
\multicolumn{2}{|p{.9\columnwidth}|}{Handwritten digits dataset consisting of 60,000 training and 10,000 testing images equally divided in 10 classes.} \\ \hline

\textbf{FashionMNIST}~\cite{fashion_mnist} & \emph{Image Recognition} \\ \hline
\multicolumn{2}{|p{.9\columnwidth}|}{Clothes recognition dataset consisting of 60,000 training and 10,000 testing images equally divided in 10 classes.} \\ \hline

\textbf{CIFAR-10}~\cite{krizhevsky2009learning} & \emph{Image Recognition} \\ \hline
\multicolumn{2}{|p{.9\columnwidth}|}{The CIFAR-10 dataset consists of 50,000 training and 10,000 testing images equally divided in 10 classes.} \\ \hline

\textbf{ImageNet}~\cite{imagenet_cvpr09} & \emph{Image Recognition} \\ \hline
\multicolumn{2}{|p{.9\columnwidth}|}{A benchmark image database organized according to the WordNet hierarchy, spanning 1,000 classes divided into 1.28 million training images, and 50,000 validation images.} \\ \hline

\textbf{Cats vs. Dogs} & \emph{Image Recognition} \\ \hline
\multicolumn{2}{|p{.9\columnwidth}|}{A subset of the ImageNet dataset consisting of 25,000 images equally divided among two classes.} \\ \hline

\textbf{GTSRB}~\cite{gstrb} & \emph{Image Recognition} \\ \hline
\multicolumn{2}{|p{.9\columnwidth}|}{The German Traffic
Sign Recognition Benchmark (GTSRB) spans 43 classes of traffic signs, divided into 39,209 training images and 12,630 test images.} \\ \hline

\textbf{WikiText}~\cite{wikitext_dataset} & \emph{Text Recognition} \\ \hline
\multicolumn{2}{|p{.9\columnwidth}|}{The WikiText subset consists of approximately 2.5 million tokens representing 720 Wikipedia articles and is divided into 2,088,628 train tokens, 217,646 validation tokens, and 245,569 testing tokens.} \\ \hline

\textbf{ESC-50}~\cite{piczak2015dataset} & \emph{Audio Recognition} \\ \hline
\multicolumn{2}{|p{.9\columnwidth}|}{The ESC-50 dataset consists of 2,000 labeled recordings equally divided between 50 classes of 40 clips per class.} \\ \hline

\end{tabular}
\end{table}

\subsection{DNN Architectures}
We evaluate \name watermark performance on different DNN architectures.
For the image classification tasks we employed three model architectures: the first architecture consists of a multilayer perceptron model composed of four fully-connected layers; the second architecture is the ResNet-18~\cite{He2016DeepRL}; the third architecture is the VGG-11~\cite{Simonyan14verydeep}. For the text classification task on \textbf{WikiText-2} dataset we employed a Transformer model composed of 2 LSTM layers
For the \textbf{ESC-50} audio classification task we employed a CNN-block-based model composed of 4 CNN layers.

\subsection{Watermarks}
We used two different watermarks of different sizes for watermarking a model using \name, namely a short \textbf{text} watermark and a large \textbf{image} watermark.
\textbf{text}-watermark corresponds to the \emph{"TATTOOED watermark!"} text sequence corresponding to 152 bits. \textbf{image}-watermark corresponds to an image of 1KB, i.e 8,192 bits.

\subsection{\name parameters}

In our experimental evaluation, we use $\gamma = 9\times10^{-4}$.
To select the $\gamma$ parameter, we performed a grid search among $\gamma$ values in the range $[1\times10^{-4}, 2\times10^{-4}, \dots, 9\times10^{-4}, 1\times10^{-3}, 2\times10^{-3}, \dots, 9\times10^{-3}, 1\times10^{-2}, 2\times10^{-2}, \dots, 9\times10^{-2}]$. We selected $\gamma = 9\times10^{-4}$ because it provides a good trade-off between fidelity and security amongst all the considered DNN architectures. In particular, with a higher $\gamma$ the security increases and fidelity decreases and vice versa when $\gamma$ decreases.
Secondly, we chose the \textit{ratio} parameter such that the number $R$ of parameters selected to embed the watermark is about 200,000 in all the considered DNN architectures to provide a unified insight on the effectiveness of \name. In real-life applications of \name, the legitimate owner can chose different parameter number ratio. This parameter, as shown in Section~\ref{sec:tattoed}, determines the length of our spreading codes. In typical CDMA applications, the spreading codes are in the order of thousands. We chose the length of our spreading codes in the order of hundreds of thousands to provide more strength to our watermark signal.

\section{Evaluation}\label{sec:evaluation}
\begin{table*}[t]
\caption{Baseline versus \name model performance comparison. }
\centering
\begin{tabular}{ l | c | c | c | c }
\hline
Model & Dataset & Baseline & \name & \name \\ 
Architecture &  & TER (\%) & TER (\%) & BER (\%) \\ \hline
ResNet & ImageNet & 24.31 & 24.34 & 0.00 \\
VGG & CIFAR10 & 15.65  & 15.70 & 0.00 \\
MLP & MNIST & 2.43  & 2.55 & 0.00 \\ 
CNN & ESC-50 & 42.23  &  42.57 & 0.00 \\ 
\hline
Model & Dataset & Baseline & \name & \name  \\ 
Architecture &  & Perplexity  & Perplexity & BER (\%) \\ \hline
Transformer & WikiText-2 & 155.59 & 154.63 & 0.00\\ \hline
\end{tabular}
\label{tab:baseline_vs_tattoed}

\end{table*}
 
We now move on to reporting the step-by-step analysis and evaluation of \name, in alignment with the watermark requirements described earlier in Section~\ref{sec:background_dnnwatermarking}.

\subsection{Fidelity}\label{sec:fidelity}
A good watermark should not deteriorate the model's performance on its original task. To this end, we compare baseline performance to the \name model's performance on the different model architectures and datasets described in~Section~\ref{sec:experimental_setup}.

Table~\ref{tab:baseline_vs_tattoed} displays the comparison between the performance of the baseline and \name watermarked models on the respective test sets. For ImageNet, CIFAR10, MNIST, and ESC-50, the performance corresponds to the test error rate (in \%), while for Wikitext-2, the performance corresponds to the perplexity. All models were trained for 60 epochs, and the watermark used is the \textbf{text}-watermark corresponding to 152 bits. In all cases, we used the $\gamma$-value equal to $9\times10^{-4}$ to embed the watermark. 
The \name model's performance shown in Table~\ref{tab:baseline_vs_tattoed} demonstrates that the use of the CDMA technique to perform watermarking does not impact the performance of the trained ML model on the intended task as the watermarked model TER is almost the same as the model's TER prior to watermarking via \name. This is due to the way CDMA works. The added signal for each watermark bit is spread over the model parameters using a low $\gamma$ value. In this way, \name does not incur significant changes to the model's weight parameters resulting in minimal impact on the model's performance on the intended task.

\subsection{Robustness}\label{sec:robustness}
A robust watermarking technique should allow the legitimate owner to correctly extract the watermark content (i.e., BER = 0) even in cases when functionality-preserving modifications are incurred on the model parameters.
For example, a common functionality-preserving modification that a DNN can undergo is fine-tuning. 
\emph{Fine-tuning} is a process that takes a model that has already been trained for one given task and then tunes or tweaks the model to make it perform better on the same task or perform a similar task. For our robustness evaluation, we consider all the possible fine-tuning techniques as follows:

\begin{figure*}[t]
    \centering        
	    \begin{subfigure}{.45\textwidth}
            \centering
            \includegraphics[width=0.8\columnwidth]{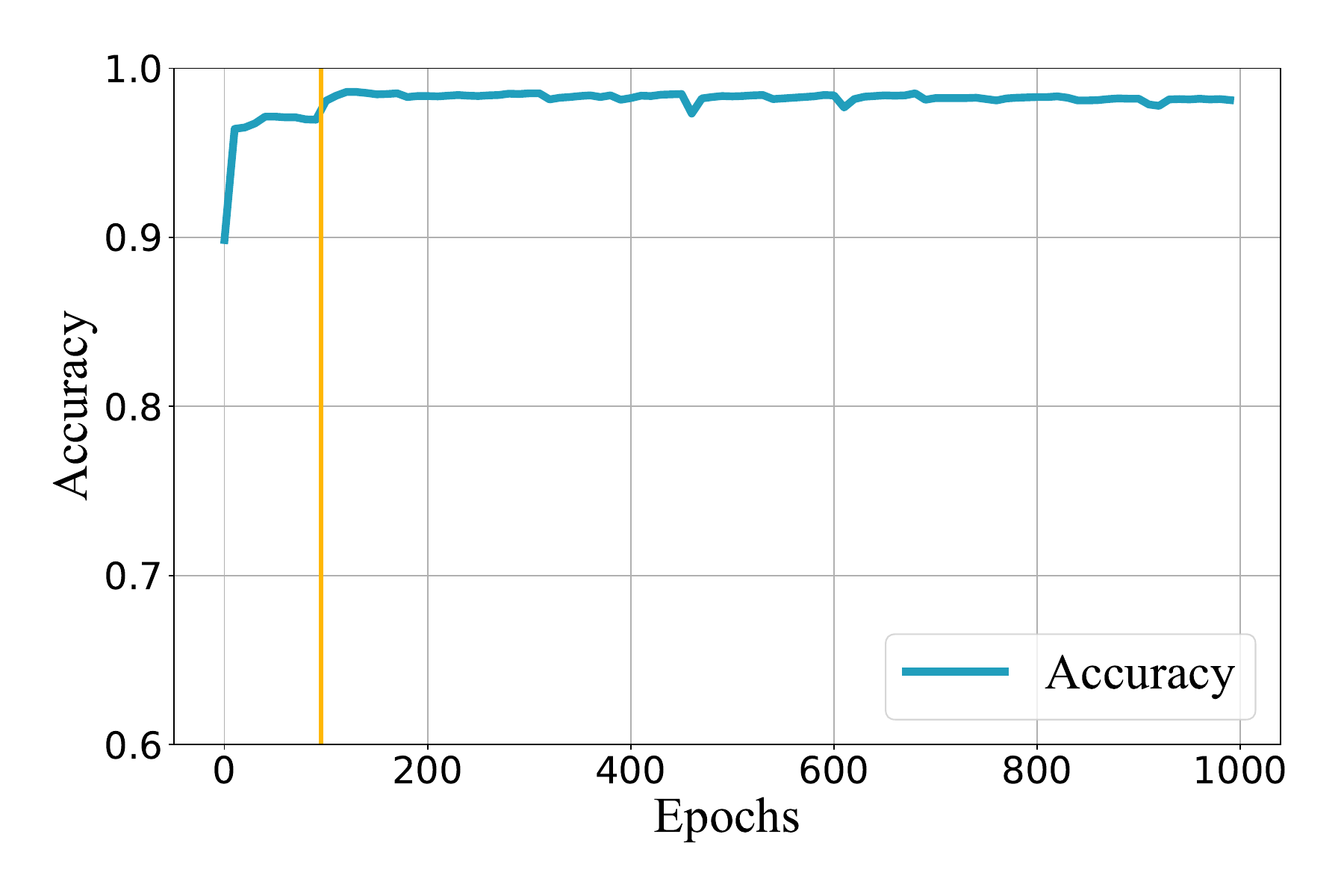}
             \caption{Multilayer perceptron model trained for 100 epochs on 30,000 MNIST instances, \name \textbf{text}-watermark embedded at epoch 100 using $\gamma = 9\times10^{-4}$, retrained (RTAL) on a different set of 30,000 MNIST instances for 1000 epochs.}
             \label{fig:retrain_mnist}
        \end{subfigure}
        \hfill
	    \begin{subfigure}{.45\textwidth}
            \centering
            \includegraphics[width=0.8\columnwidth]{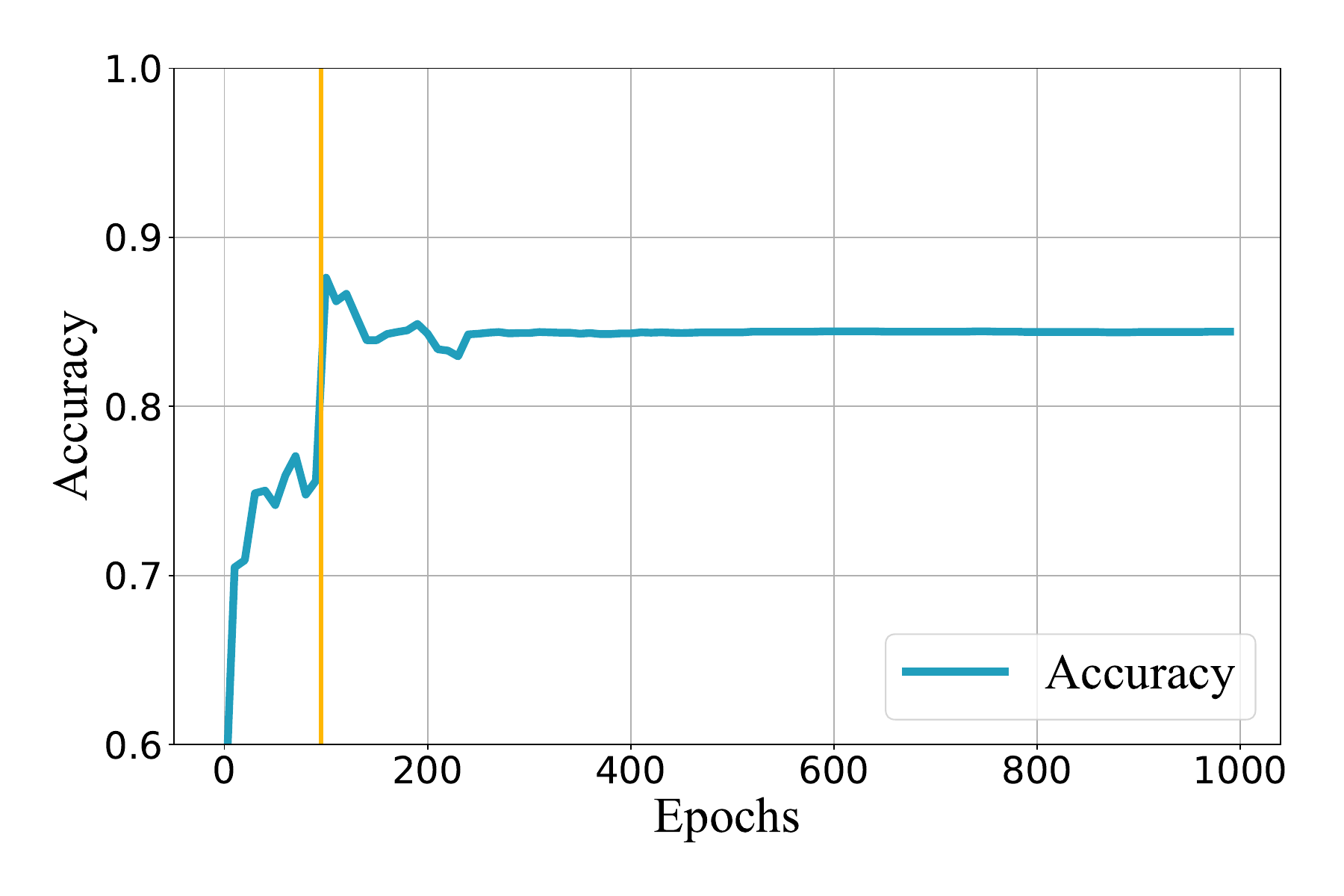}
             \caption{VGG-11 model trained for 100 epochs on 30,000 CIFAR10 instances, \name \textbf{image}-watermark embedded at epoch 100 using $\gamma = 9\times10^{-4}$, retrained (RTAL) on a different set of 30,000 CIFAR10 instances for 1000 epochs.}
             \label{fig:retrain_cifar}
        \end{subfigure}
        \bigskip
    \vspace{-1em}
    \caption{The effect of RTAL on \name watermark using the same amount of additional data to retrain the model as that used for initial training. In each case, the watermark BER is equal to 0.} 
    \label{fig:retrain}
\end{figure*}

\leavevmode \\
\textbf{Retrain all layers (RTAL).} RTAL is a method that retrains the DNN model on a dataset of the same size as the one used in training.
To evaluate the robustness of \name watermarks under RTAL, we split MNIST and CIFAR10 datasets into two parts, one for the initial training and one for retraining. Each of these splits contained 30,000 images. We trained the multilayer perceptron model on the first half of the MNIST and the VGG-11 model on the CIFAR10 dataset for 100 epochs and then used \name to watermark each model. After that, we used the second half of the dataset to fine-tune the models for 1,000 epochs.
Figure~\ref{fig:retrain_mnist} shows the validation accuracy of the model trained on the MNIST task. We watermarked the model at epoch 100 (yellow vertical line) with the \textbf{text} watermark. During the retraining phase, the validation accuracy of the model increases because it is being trained on new data. After every epoch, we checked if the watermark was present in the model and confirmed that the watermark was unaffected even after 1,000 epochs.
Figure~\ref{fig:retrain_cifar} displays the validation accuracy on the CIFAR10 task. Similar to MNIST, we trained the model on the first half of the CIFAR10 dataset for 100 epochs and then watermarked it with the \textbf{image} watermark. Afterward, we retrained the model using the second half of the dataset for 1,000 epochs. After every epoch, we checked if the watermark was present in the model and confirmed that the watermark was unaffected even after 1,000 epochs.

\leavevmode \\
\textbf{Fine-tune all layers (FTAL).}
Fine-tune all layer is a common fine-tuning technique to boost the performance of a pre-trained model on the task using new data.
To evaluate the robustness of \name after an FTAL procedure, we performed two experiments: with MNIST and with CIFAR10. First, we trained the multi-layer perceptron model on 40,000 MNIST instances for 100 epochs, watermarked it using the \textbf{text}-watermark with $\gamma = 9\times10^{-4}$ and then performed FTAL using the remaining 10,000 MNIST instances for another 100 epochs. 
Then, we trained the VGG-11 model on 40,000 CIFAR10 instances for 100 epochs, watermarked it using the \textbf{image}-watermark with $\gamma = 9\times10^{-4}$ and then performed FTAL using the remaining 10,000 CIFAR10 instances for another 100 epochs. 

\leavevmode \\
\textbf{FTAL using a different Dataset.}
\begin{figure*}[t]
    \centering        
        \begin{subfigure}{.45\textwidth}
            \centering
            \includegraphics[width=0.8\columnwidth]{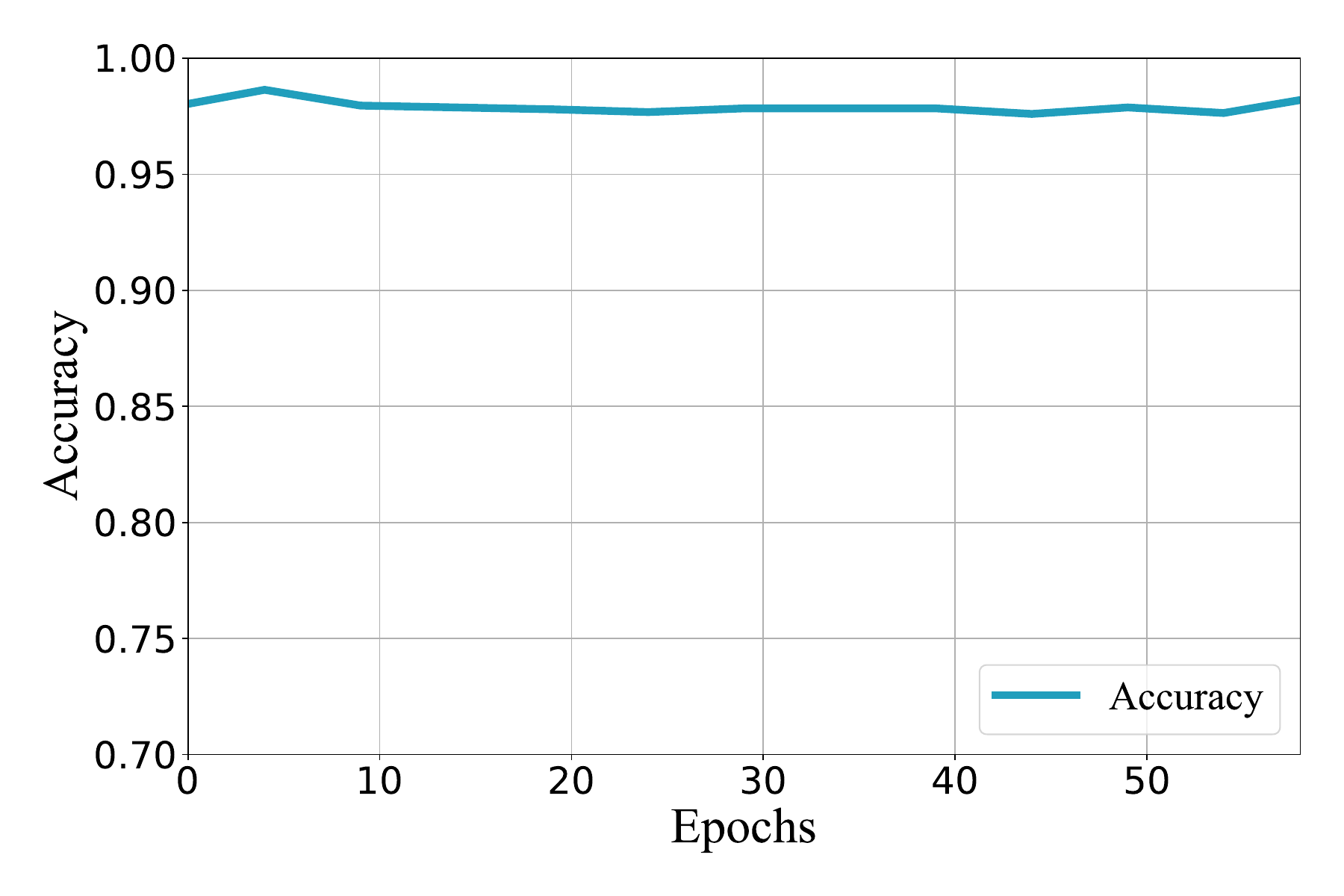}
             \caption{ResNet-18 model pre-trained on ImageNet, \name \textbf{image}-watermark embedded using $\gamma = 9\times10^{-4}$, fine-tuned on Cats vs. Dogs for 60 epochs.}
             \label{fig:fine_tune_imagenet_pets}
        \end{subfigure}
        \hfill
	    \begin{subfigure}{.45\textwidth}
            \centering
            \includegraphics[width=0.8\columnwidth]{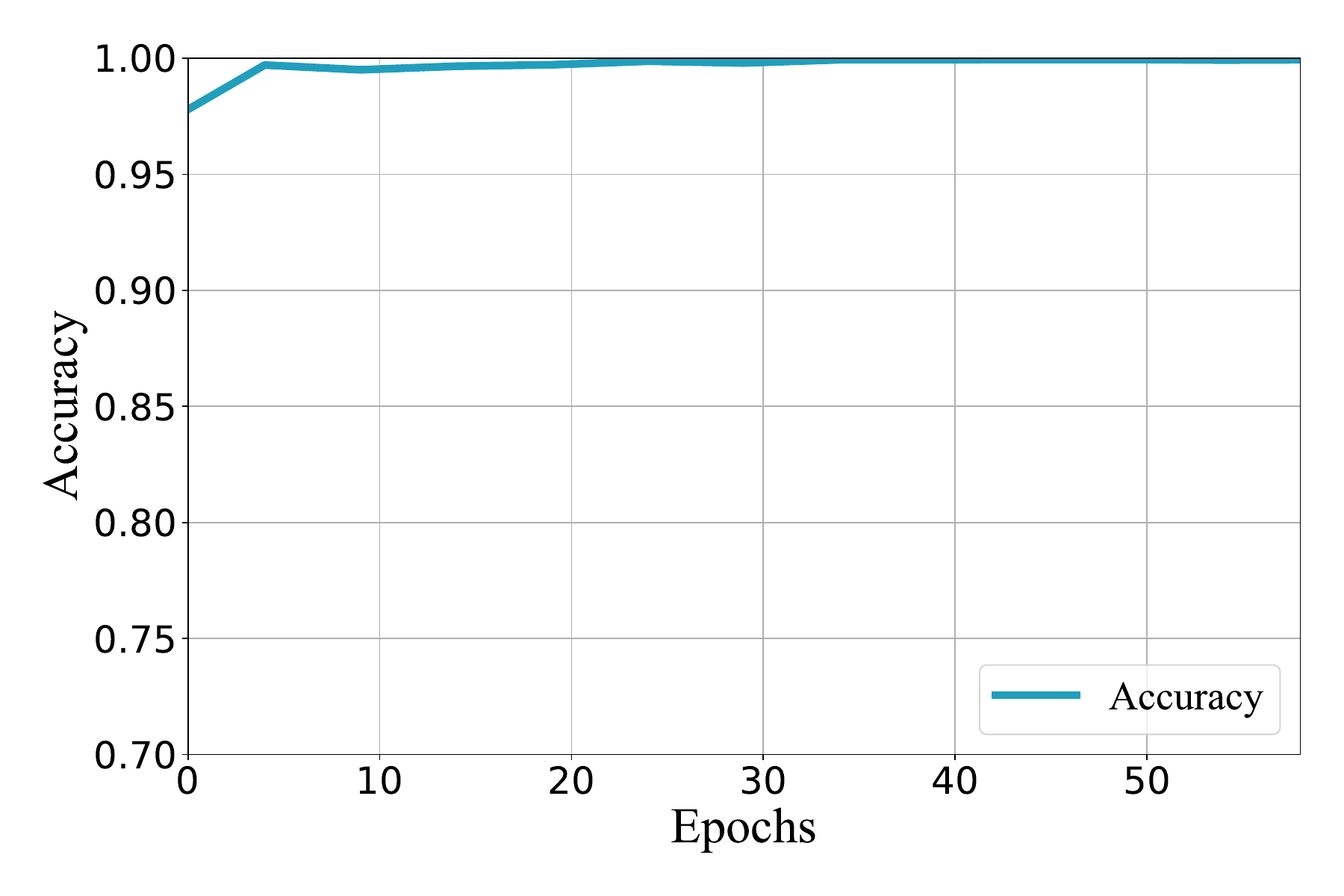}
             \caption{VGG-11 model pre-trained on CIFAR10, \name \textbf{image}-watermark embedded using $\gamma = 9\times10^{-4}$, fine-tuned on GTSRB for 60 epochs.}
             \label{fig:fine_tune_cifar_gtsrb}
        \end{subfigure}
        \bigskip
    \vspace{-1.2em}
    \caption{The effect of model fine-tuning on \name watermark using a different dataset than the one used for the training. In each case, after the fine-tuning procedure, the watermark BER was equal to 0.} 
    \label{fig:fine_tune_different_dataset}
\end{figure*}
FTAL using a different dataset is a fine-tuning technique that serves to re-purpose a model on a new similar task.
To evaluate the robustness of \name after such FTAL, we perform three experiments:
\begin{itemize}
    \item \textbf{ImageNet repurposed on Cats vs. Dogs}: we trained the ResNet-18 architecture on the ImageNet dataset and watermarked it with the \textbf{image}-watermark using \name and $\gamma = 9\times10^{-4}$. We then fine-tuned this model to solve the Cats vs. Dogs task for 60 epochs. 
    \item \textbf{CIFAR10 repurposed on GTSRB}: we trained the VGG-11 architecture on the CIFAR10 dataset and watermarked it with the \textbf{image}-watermark using \name and $\gamma = 9\times10^{-4}$. We then fine-tuned this model to solve the GTSRB task for 60 epochs. 
\end{itemize}
Figure~\ref{fig:fine_tune_different_dataset} displays the model accuracy in each epoch during the fine-tuning procedure. As we can see, each model performs well in the new task, thus showing that the fine-tuning procedure is performing as expected.

\textbf{Results.}
After each of the abovementioned procedures we extracted the watermark to check its presence in the model. In each case, the watermark presence was successfully verified with a watermark BER of 0, showing that the RTAL, FTAL and FTAL with a different dataset technique does not hamper the \name watermark. From this evaluation, we can see that FTAL, RTAL, and FTAL with different datasets do not disrupt the \name watermark, thus satisfying the \textbf{Robustness} requirement.

\subsection{Security}\label{sec:security}
A secure watermarking scheme ensures that the legitimate owner of a model can be verified with high confidence, and no malicious entity can remove or overwrite the watermark and claim ownership of the model. A DNN watermarking algorithm is secure if the malicious entity to destroy the watermark: a) either renders the model unusable in its intended task or b) requires a cost, in terms of resources (e.g., data, time, computational power), greater than the cost that the legitimate owner incurred to produce the watermarked model.
In our evaluation we consider three state-of-the-art watermark removal techniques namely REFIT~\cite{chen_refit} and parameter pruning~\cite{neural_cleanse,li2021neural}.

\leavevmode \\
\textbf{REFIT~\cite{chen_refit}}
To evaluate the security of the \name watermark, we tested it against REFIT~\cite{chen_refit}, a state-of-the-art watermark removal technique. REFIT is agnostic to the technique used to watermark a DNN model, and can remove a portion of watermark sufficient to make a wide range of prior neural network watermarking schemes unreliable, such as~\cite{uchida_embedding,wang_dnn2020,song2017machine,deepsign2019,deepmark_2019,backdoor2018watermark,Chen2019BlackMarksBM,Jia2021EntangledWA,Merrer2017AdversarialFS}.
REFIT proposes four separate watermark removal techniques. The first technique relies on a fine-tuning procedure with a maliciously configured learning rate schedule, the second relies on unlabeled data augmentation (AU), the third relies on elastic weight consolidation (EWC), and the fourth simultaneously employs AU and EWC to remove the watermark.
To test the security of the watermark \name against the REFIT attack, after training our models and watermarking them using \name, we employed the REFIT removal techniques for the same number of epochs used in the original training.
Typically, fewer epochs are used during the REFIT attack; otherwise, the adversary attempting to remove the watermark would have to spend the same amount of computation resources needed to train the DNN model from scratch. We used the original REFIT implementation in our experiments.\footnote{\url{https://github.com/sunblaze-ucb/REFIT}}

\textbf{Results.}
We trained the VGG-11 model on the CIFAR10 dataset for 60 epochs and used \name to watermark the model with the \textbf{text}-watermark using a $\gamma = 9\times10^{-4}$. We pitch \name against each of the four watermark removal techniques proposed by REFIT using the parameters suggested in the published paper~\cite{chen_refit}.
For each REFIT removal attack, after each epoch we extract the \name watermark to check whether the watermark could be correctly extracted from the network. In every case, the watermark was extracted with a watermark BER of 0, showing that REFIT~\cite{chen_refit} watermark removal techniques are ineffective in removing the \name watermark from the model.
\begin{table}[t]
\caption{REFIT~\cite{chen_refit} watermark-removal technique impact on prior works and \name.}
\centering
\begin{tabular}{l cccc}
\toprule
 & \multicolumn{4}{c}{REFIT~\cite{chen_refit}}\\
{Method} & Basic & AU & EWC & EWC+AU \\ 
\cmidrule(r){1-1} \cmidrule(l){2-5}
Uchida et al.~\cite{uchida_embedding} & \xmark & \xmark &  \xmark & \xmark \\ 
Zhang et al.~\cite{zhang2018protecting}  & \xmark & \xmark &  \xmark & \xmark \\ 
Adi et al.~\cite{backdoor2018watermark}  & \xmark & \xmark &  \xmark & \xmark \\ 
Namba et al.~\cite{namba2019robust}  & \xmark & \xmark &  \xmark & \xmark \\ 
Wang et al.~\cite{wang_riga}  & \xmark & \xmark &  \xmark & \xmark \\ 
\name   & \cmark & \cmark &  \cmark & \cmark \\ 
\bottomrule
\end{tabular}
\label{tab:refit_comparisons}
\vspace{-1em}
\end{table}
To highlight that \name is pushing forward the state-of-the-art in DNN watermarking, in Table~\ref{tab:refit_comparisons} we compare the impact that REFIT has on prior art and \name. 
Following REFIT~\cite{chen_refit}, in Table~\ref{tab:refit_comparisons}, we display the performance of REFIT on the representative works~\cite{uchida_embedding,zhang2018protecting,backdoor2018watermark,namba2019robust,wang_riga} in the different watermarking embedding schemes. For~\cite{zhang2018protecting,backdoor2018watermark,namba2019robust} we report the results from REFIT~\cite{chen_refit}. As we can see, REFIT can disrupt the watermark verification of those techniques. Wang et al.~\cite{wang_riga} report that REFIT can disrupt the watermark, thus hindering the verification process, but the REFIT parameters to fully remove the watermark from the model are difficult to find.
On the other side, \name watermark is completely unaffected by REFIT, and the watermark BER is always 0, leading to an improvement over the prior art in terms of security. 

\leavevmode \\
\textbf{Parameter Pruning.}
\emph{Parameter pruning} is a technique that removes neurons from an already trained neural network while preserving the model's performance on the intended task. 
Typically, parameter pruning is performed by zeroing the weight parameters of the model.
To evaluate \name watermark against the parameter pruning technique, we prune the parameters of the watermarked model by varying the amount of pruning from 25\% to 99.99\%.

Table~\ref{table:result_pruning} displays the effects that different amounts of pruning have on the \name watermark BER and the TER of the VGG-11 model trained on the CIFAR10 dataset.
Table~\ref{table:result_pruning} shows that after pruning up to 99\% of the model parameters, the \name watermark BER is 0. However, the TER goes to 90\%, which is no better than guessing among the 10 classes of CIFAR10; thus, the model is unable to perform its intended task.
The inability of parameter pruning to remove the \name watermark, even when 99\% of the parameters are pruned, is due to the foundations of the \name watermarking technique. \name employs CDMA spread-spectrum channel coding to embed the watermark in a randomly selected portion of the model's parameters. As mentioned in Section~\ref{sec:tattoed}, to extract a bit of the watermark $m_i$, the spreading code is multiplied by the vector of the parameters that were selected for watermarking.
Even if the random pruning can zero 99\% of them, the remaining non-zero weights contribute enough to correctly decoding $m_i$ to allow the detection of the watermark in the model. This ability derives from the inherent capabilities of the CDMA, where the adversary cannot prevent the communication without completely destroying the channel (thus zeroing all the parameters of the model in our case).  

\leavevmode \\
\textbf{Parameter Shuffling}
An adversary might attempt to reorder the parameters in each network layer to prevent the watermark verification. Firstly, the reordering of the parameters should be carefully done to not impact the model performance. Secondly we can find and reverse back to the original order the parameters of the networks as follows.
Let $W = \{W_1, W_2, \ldots, W_k\}$ be our original watermarked neural network. $W’ = \{W’_1, W’_2, \ldots, W’_k\}$ represents the resulting network after the adversary shuffles the parameters. To undo the shuffling we need to compute the cosine distance between each layer of the network (i.e., submatrix).
For simplicity, lets consider the recovery process of $W_1$, from the shuffled version $W_1^{'}$. To compute the cosine distance we first need to compute the vector norm for $W_1$ and $W_1^{'}$:
\begin{table}[t]
\caption{Parameter Pruning watermark removal technique on VGG-11 model and CIFAR10 task trained for 100 epochs and watermarked with \name using the \textbf{text}-watermark. }
\centering
\begin{tabular}{lrr}
\toprule
Pruning &TER (\%) &BER (\%) \\
\midrule
$25.00\%$ & $15.68$ & $0.00$  \\
$50.00\%$ & $19.34$ & $0.00$  \\
$75.00\%$ & $90.02$ & $0.00$  \\
$90.00\%$ & $90.00$ & $0.00$  \\
$95.00\%$ & $90.00$ & $0.00$  \\
$99.00\%$ & $90.00$ & $0.00$  \\
$99.75\%$ & $90.00$ & $9.68$  \\
$99.99\%$ & $90.00$ & $100.00$  \\
\bottomrule
\end{tabular}
\label{table:result_pruning}
\end{table}
\begin{equation}
    Norm_{W_1} = \sqrt{Diag(W_1^{\top} W_1)}
\end{equation}

\begin{equation}
    Norm_{W_1^{'}} = \sqrt{Diag(W_1^{'\top} W_1^{'})}
\end{equation}

Where $T$ represents the transpose of the matrices and $Diag$ means that we keep the diagonal values of the resulting matrix after the matrix multiplication.
Assuming that $Norm_{W_1}$ and $Norm_{W_1^{'}}$ are column matrices, then we can compute the cosine distance like this:

\begin{equation}
    CosDist(W_1 , W_1^{'}) = W_1^{\top} W_1^{'} \oslash Norm_{W_1} Norm_{W_1^{'}}^\top
    \label{eq:cosdist}
\end{equation}

where $\oslash$ represents the element wise division.
The cosine matrix~\ref{eq:cosdist} is a squared matrix, in which the rows indicate the original network order for the neurons in the first hidden layer, and the columns indicate the order of the neurons in the new network. If the network has not been shuffled, the diagonal elements of this matrix will be one. Otherwise, the ones in each row indicate the performed shuffling for that neuron. For example, if the entry $(1,3)$ in the cosine matrix~\ref{eq:cosdist} would be one, it means that the first neuron in the original network has been moved to the third position in the shuffled network. There is only one one per column and row in that matrix. Knowing the shuffled elements, we can undo the shuffling in this layer and perform the same process for every other layer in the network, recursively.After unshuffling every layer, we can recover the watermark using the \textbf{Extract} algorithm~\ref{alg:verify_watermark}.

\leavevmode \\
\textbf{Watermark Overwriting.} Another type of attack against the watermark is overwriting, where an adversary that knows the watermarking technique used attempts to insert another watermark in the model in an attempt to disrupt the original watermark embedded by the legitimate owner. In \name, this is handled by design from its reliance on CDMA. CDMA allows multiple users to simultaneously embed different signals by using different spreading codes. Given that the spreading codes used by the legitimate owner are secret (and cannot be guessed), this ensures that any overwriting attack cannot inhibit the legitimate owner from verifying the presence of his/her original watermark. Even though the robustness towards overwriting attacks is handled by design, we also performed several experiments to confirm our claims.  We attempted to embed a different watermark using the \name technique (with different spreading codes) in a previously watermarked model. This did not impact the ability to correctly extract the initial (i.e., legitimate) watermark from the model. 

\subsection{Integrity}\label{sec:integrity}
As explained in Section~\ref{sec:background_dnnwatermarking}, for a DNN watermarking scheme to satisfy the \textit{integrity} property, its \textit{Extract} method should, in the absence of modifications to the watermarked model, exactly output the previously-inserted watermark $m$. As done in prior work~\cite{barni_survey_wtm}, we use BER to evaluate the \textit{integrity} of \name watermark. 
In Table~\ref{tab:baseline_vs_tattoed} we display the \name watermark BER extracted without any modification to the watermarked model. We perform this analysis on multiple domains and architectures. In all cases, the BER of the extracted watermark was always 0, thus showing that \name satisfies the \textit{integrity} property.

\subsection{Efficiency}\label{sec:efficiency}

\begin{table}[t]
\caption{\textbf{Mark} and \textbf{Extract} algorithms performance on the multi-layer perceptron architecture with 200,000 weight parameters and watermarked via \name using the \textbf{text}-watermark with a $\gamma=9\times10^{-4}$.}
\centering
\begin{tabular}{ c c c }
\hline
Ratio & Time to Mark & Time to Extract \\ \hline
$12.50\%$ & 1.36sec & 0.42sec \\
$25.00\%$ & 2.73sec  & 0.67sec\\ 
$50.00\%$ & 5.35sec & 0.94sec\\
$100.00\%$ & 11.14sec  &  1.90sec\\ \hline
\end{tabular}
\label{tab:efficiency}
\end{table}

\name watermarking technique is a one-pass post-training procedure. For this reason, to evaluate \name performance we measured the time (in seconds) that \name requires to embed and verify the watermark from a neural network containing about 200K parameters while embedding the watermark using different portions of the network starting from $12.5\%$.
The results in Table~\ref{tab:efficiency} show that \name needs less than 12 seconds to watermark a model with about 200K parameters and less than 2 seconds to verify whether the watermark is present in the model.
Table~\ref{tab:efficiency} demonstrates that the computational complexity of \name grows linearly with the number of parameters chosen to embed the watermark.
Differently, prior DNN watermarking techniques such as~\cite{uchida_embedding,backdoor2018watermark,li_paper_wtm}, due to the model modifications they perform to watermark a model, require multiple training epochs (e.g., an epoch on MNIST requires around 10 seconds on a high-end desktop GPU). 

\subsection{Generality}\label{sec:generality}
The \emph{generality} property requires the watermarking technique not to be restricted to particular neural network architecture or dataset to watermark a DNN model correctly.
We empirically verified the \emph{generality} property of the \name watermark by employing it on six image classification tasks (MNIST, FashionMNIST, CIFAR10, GTSRB, ImageNet, Cats vs. Dogs), one audio classification task (ESC-50), and one language modeling task (Wikitext-2), and several model architectures (multi-layer perceptron, CNN-based (ResNet18, VGG11) and LSTM-based). In each of the considered tasks, the \name watermarking technique could be applied seamlessly without requiring domain or architecture-specific modifications thus satisfying the watermark \textbf{generality}.

\section{Related Work}\label{sec:related_work}

\subsection{White-Box Watermarking Techniques}

The work by Uchida et al.~\cite{uchida_embedding,Nagai2018DigitalWF} is the first attempt to watermark a DNN. The embedding is made possible by using an extra regularization term in the cost function that introduces a statistical bias on the weight parameters of the learned ML model. This statistical bias is then used to infer information about the ownership of the ML model.
Li et al.~\cite{li_paper_wtm} extend the work of Uchida et al.~\cite{uchida_embedding,Nagai2018DigitalWF} by introducing an extra regularization term based on informed coding. This allows the author to insert a bigger watermark with a lower impact on the model accuracy, thus improving the capacity of their scheme compared to previous work. However, their usage of informed coding as a regularization term has been proved less robust against removal techniques, such as FTAL with a different dataset, compared to the original one proposed by Uchida et al., as reported in table 10 of work~\cite{li_paper_wtm}.
Wang et al.~\cite{wang_dnn2020} insert an independent neural network that allows the use of selective weights to watermark another ML model. This independent model is kept secret and serves as the watermark verification tool. The watermark is embedded by training the two neural networks simultaneously. In this procedure, the independent neural network uses selected target parameter weights of the model to be watermarked to embed the watermark information. To extract the watermark, the input layer of the independent neural network is connected to the selective weights of the marked neural network, and the watermark verification is done by observing the outputs of the independent neural network.
Song et al.~\cite{song2017machine} present a watermarking technique with the goal of embedding information related to the private training dataset into the trained model's weight parameters. They achieve this by either encoding sensitive information about the private training dataset in the least significant bits and the signs of the model's weight parameters.
DeepSigns~\cite{deepsign2019} works by embedding the watermark in the probability density function of the data abstraction obtained by a DNN layer. The watermark verification is done by using a subset of training data, querying the model with them, and computing the statistical mean value of the activation features obtained by passing these input data in the model. The acquired mean values are used to extract the watermark content.

Along the same line of thought as Uchida et al.~\cite{uchida_embedding,Nagai2018DigitalWF}, Wang et al.~\cite{wang_riga}, propose a GAN-based adversarial training technique to embed the watermark in a manner that encourages the weights distribution of a watermarked model to be similar to the weights distribution of a non-watermark model. 

\subsection{Threats to Watermarking Techniques}
Several works have been proposed to assess watermark resilience to malicious adversaries trying to remove~\cite{Hitaj2019evasion,wang2019attacks,namba2019robust,neural_cleanse}, overwrite~\cite{wang2019attacks}, forge~\cite{Xu2019ANM,fan2019rethinking} and detect them~\cite{wang2019attacks,shafieinejad_watermarking_robustness}. 
Wang et al.~\cite{neural_cleanse} propose a generalized technique to detect and mitigate backdoors in DNNs that can be applied to mitigate backdoor-based watermarking schemes.
Chen et al.~\cite{chen_refit} propose REFIT, a framework to remove watermarks through a type of malicious modification of the watermarked model. They show that making a model forget part of the task it was initially trained on can effectively remove watermarks. They do that by devising a malicious learning rate schedule, using additional data, and using elastic weight consolidation~\cite{Kirkpatrick2017OvercomingCF}. REFIT was shown to be effective in removing watermarks based on prior techniques~\cite{uchida_embedding,Nagai2018DigitalWF,wang_dnn2020,song2017machine,deepsign2019,backdoor2018watermark,Chen2019BlackMarksBM,wang_riga}.
On the contrary, as we show in Section~\ref{sec:security}, \name is unaffected by the removal techniques proposed by REFIT. 

Xu et al.~\cite{Xu2019ANM} propose an attack that can forge the backdoor-based watermark by finding the trigger inputs and then claiming ownership of the model. \name does not use any trigger inputs, thus making this type of attack ineffective. Moreover, due to \name foundations on CDMA, only the proprietary of the spreading codes (i.e., the secret key to generate them) can retrieve the watermark and claim ownership, thus rendering a forging attack such as~\cite{Xu2019ANM} unfeasible. 
\section{Discussion}\label{sec:limitations}

Over the years, various approaches have attempted to watermark neural networks by performing different types of weight modifications, such as modifying the least significant bits or introducing a bias in the distribution of the weight parameters of the trained network~\cite{uchida_embedding,wang_dnn2020,song2017machine,deepsign2019}. 
\name relies on spread spectrum channel coding to watermark a model by directly coding the watermark content (bits) in the weight parameters of the model. The employed channel coding technique allowed \name to watermark a model and surpass prior DNN watermarking techniques in terms of the security and robustness it offers toward watermark removal attempts. 

Nevertheless, there are inherent limitations to current DNN watermarking approaches, that can also affect \name. For completeness, this section discusses potential threat vectors that fall outside the threat model typically considered in white-box watermarking literature.

In section~\ref{sec:evaluation}, we showed that \name is robust towards parameter pruning when pruning is done by zeroing the weight parameters of the network. Another way to perform parameter pruning is to physically remove the pruned neurons from the model, resulting in a new model architecture. This technique is known as model compression~\cite{model_compression}. Given that \name relies on an ordering of the weight parameters to embed and extract the watermark (see section~\ref{sec:tattoed}), a completely new architecture that results after pruning would impact the correct extraction of the watermark from the newly obtained model.
To thwart such compression-based attacks, the legitimate owner can compress the model prior to using \name for watermarking to provide another level of security to the watermark since, once compressed, an ML model can not be reduced by a large amount without impairing performance. Moreover, performing model compression requires significant domain knowledge and resources, rendering this threat less likely to be applied in practice.

Another attack vector mentioned in section~\ref{sec:evaluation} is model parameter shuffling, where the adversary shuffles the weight parameters of the stolen model to avoid detection. We showed that we can tackle this kind of attack by first performing an un-shuffling phase prior to watermark verification. Nevertheless, the adversary may go one step further, and besides shuffling the parameters, he can also introduce new neurons (or layers), changing the structure of the model. In this case, we can not perform the un-shuffling procedure. However, the legitimate owner can compare the suspicious model architecture to identify the newly introduced neurons by the adversary, remove them, and then perform the un-shuffling procedure. 
Even though parameter shuffling combined with introducing new parameters can limit \name watermark verification, it also requires significant domain knowledge by the adversary. Shuffling has to be carefully done not to deteriorate the performance of the model, and introducing new neurons has to be carefully done for the same reason, and in some cases, would require a fine-tuning step requiring the adversary to have both the computational power and a high-quality dataset to carry out the procedure, thus rendering this attack unlikely in practice.

One approach shown to be highly effective in thwarting DNN watermarking techniques is model distillation~\cite{modeldistillation}. Model distillation is a technique in machine learning where a smaller model, commonly referred to as the student, is trained to replicate the performance of a larger model, referred to as the teacher. This is achieved by having the student model learn from the teacher model's predictions rather than the original training data, aiming to build a more efficient model in terms of computational resources while exhibiting a similar performance to the teacher model.
In an attempt to remove the \name-watermark (or any other type of DNN watermark), the adversary can choose to construct a new model by relying on model distillation. This technique would result in a completely new model that will not contain the watermark. However,
the distillation procedure is highly complex and computationally intensive, and it has been shown that the student model is hard to train and usually performs worse than the teacher model, rendering it a sub-optimal choice for the adversary. Concluding, cybersecurity approaches are rarely insurmountable for adversaries. Rather, the goal is to increase the challenge for adversaries. This is our aim with \name. We demonstrate that \name 
surpasses prior art in the watermarking domain, while making it considerably harder for an adversary to remove a watermark from a DNN model. Finally, we stress that the limitations discussed in this section apply to \textit{all} existing watermarking approaches.

\section{Conclusions}\label{sec:conclusions}
This paper introduced \name, a novel white-box neural network watermarking technique based on CDMA spread-spectrum channel-coding. 
Our extensive evaluation showed that \name watermark incurs no penalty on the model performance, and it can be applied to a broad class of model architectures and tasks. We demonstrated that \name watermark remains unaffected by state-of-the-art watermark removal techniques, thus improving the state-of-the-art white-box watermarking of neural networks.
Our evidence, reported in this paper and supported by our comprehensive evaluation, indicates that \name is more robust and secure against watermark removal attempts than prior work in this domain. The generality of \name coupled with its ability to ensure robustness and security, even in extreme adversarial settings, make \name a strategy of choice for DNN IP protection.

\section{Acknowledgments}
This work was partially supported by project SERICS (PE00000014) under the MUR National Recovery and Resilience Plan funded by the European Union - NextGenerationEU.

\bibliographystyle{acm} 
\bibliography{bibliography}

\end{document}